\documentclass[letter]{aa} 

\usepackage{graphicx}
\usepackage{txfonts}
 \usepackage[breaklinks=true]{hyperref} 
 \usepackage{natbib,twoopt}
 \usepackage{float}
 \usepackage[breaklinks=true]{hyperref} 
 \bibpunct{(}{)}{;}{a}{}{,}    

\begin{document}

\title{The inner environment of Z~CMa: High-Contrast Imaging Polarimetry with NaCo
    \thanks{Based on observations made with the \textit{VLT}, program 094.C-0416(A).}
}

\author{H. Canovas\inst{1,6}, S. Perez\inst{2,6}
   C. Dougados\inst{3},
   J. de Boer\inst{4,7},
   F. M\'enard\inst{3},
   S. Casassus\inst{2,6},
   M.~R. Schreiber\inst{1,6},
   L.~A. Cieza\inst{5,6},
   C. Caceres\inst{1,6},
   J.~H. Girard\inst{7}
}

\institute{ 
Departamento de F\'isica y Astronom\'ia, Universidad de Valpara\'iso,
Valpara\'iso, Chile, \email{hector.canovas@dfa.uv.cl} 
\and 
Departamento de Astronom\'ia, Universidad de Chile, Casilla 36-D,
Santiago, Chile 
\and
   UMI-FCA, CNRS/INSU, France (UMI 3386), and Dept. de Astronom\'{\i}a, Universidad de Chile, Santiago, Chile. 
\and
Sterrewacht Leiden, Universiteit Leiden, P.O. Box 9513, 2300 RA Leiden, The Netherlands.
\and
Facultad de Ingenier\'ia, Universidad Diego Portales, Av. Ejercito 441, Santiago, Chile
\and
  Millennium Nucleus ``Protoplanetary Disks in ALMA Early Science"
\and
European Southern Observatory, Casilla 19001, Santiago, Chile.
}

\date{\today}

\abstract
{ Z\,CMa is a binary composed of an embedded Herbig Be
  and an FU Ori class star separated by $\sim100$ au. Observational evidence
  indicate a complex environment in which each star has a circumstellar
  disk and drives a jet, and the whole system is embedded in a large dusty
  envelope.}
{We aim to probe the circumbinary environment of Z\,CMa in the inner 400 au in scattered light.}
{ We use high contrast imaging polarimetry with VLT/NaCo at $H$ and $K_s$ bands.}
{The central binary is resolved in both bands. The polarized images
 show three bright and complex structures: a common dust envelope, a sharp extended feature
 previously reported in direct light, and an intriguing bright clump located $0\farcs3$ south of the binary, which
 appears spatially connected to the sharp extended feature.}
{We detect orbital motion when compared to previous observations, and report a new outburst driven by the Herbig star.
Our observations reveal the complex inner environment of Z\,CMa with unprecedented detail and contrast. 
}


\titlerunning{The inner environment of Z~CMa: High-Contrast Imaging Polarimetry with NaCo}
\authorrunning{Canovas et al.}

\maketitle


\section{Introduction}

Herbig stars are the earliest optically-revealed stage of medium mass star
formation. They are usually found embedded in envelopes rich in gas and dust,
typically surrounded by circumstellar protoplanetary disks. Stellar multiplicity
is an important element which has a strong impact on circumstellar evolution.
Interestingly, the majority of stars are born in multiple systems \citep{2012ApJ...745...19K}.
Z~Canis~Majoris (Z\,CMa) is an enigmatic binary system composed of an
embedded Herbig Be star (primary) and an FU Ori (secondary) star
separated by $\sim100$ au \citep{1991AJ....102.2073K}. 
Distance estimations to Z\,CMa range from 930 to 1150 pc
\citep{1974A&A....37..229C,2000MNRAS.312..753K}.
The whole system is surrounded by a huge infalling envelope
\citep{1991ApJ...381..250B, 2009A&A...497..117A}, which can explain
the very high accretion rate of the FU Ori star
\citep[$\sim3\,\times\,10^{-5}\,\mathrm{M_{\sun}\,yr^{-1}}$,][]{2004MNRAS.349.1516V}.
Spectropolarimetric observations of Z\,CMa revealed the nature of the primary
showing a spectrum rich in emission lines and largely polarized \citep{1993ApJ...417..687W}.
A dusty envelope or cocoon around the primary is most likely causing this effect.
The extinction \citep[$A_v\sim10\,(5.2)$ for the primary (secondary),][]{2013ApJ...763L...9H}
created by this cocoon can decrease the brightness of the primary
towards optical wavelengths \citep[see also ][]{1993ApJ...402..605W, 2010A&A...509L...7S}.
The primary drives a 3.6 pc jet \citep{1989A&A...224L..13P}, and IFU
observations with OSIRIS/Keck show that the secondary drives a ``microjet''
\citep[$\sim400$ au,][]{2010ApJ...720L.119W}.
At a few hundred au south of Z~CMa, \citet{2002ApJ...580L.167M}
reported the presence of a jet-like feature observed in scattered light extending about
$\sim1000$~au from Z\,CMa towards the southwest. It is not clear whether it is associated
with a bipolar outflow, or a jet, or whether it is a completely independent structure.

Z\,CMa's light curve shows strong and moderate outbursts with
variations both in duration and in strength. During quiescence the secondary
dominates the optical and near-infrared (IR) up to $H$ band and the primary
dominates at longer wavelengths. However, during the strong outburst episodes
the Herbig star becomes the dominant source at all wavelengths
\citep{2004MNRAS.349.1516V, 2013ApJ...763L...9H}.
Between 2008 and 2009 Z\,CMa suffered a strong outburst ($\Delta m_v \approx 2.5$).
While some observational evidence favors an accretion burst from the primary
as the source of this outburst \citep{2010A&amp;A...517L...3B, 2013ApJ...763L...9H},
other observations favor the formation of holes in the dust cocoon around the primary
to explain the sudden increment in flux \citep{2010A&A...509L...7S}. 
Changes in the structure and/or optical thickness of the dust cocoon can
naturally explain the outburst: the opening of new holes in the direction
of the line of sight allows the light from the Herbig star to directly escape,
resulting in a dramatic increment of the total flux of the system at optical
and NIR wavelengths. \citet{2012A&amp;A...543A..70C}
presented polarized differential images (PDI) at optical wavelengths,
showing evidence of light escaping through a hole in the dust cocoon and cavities
carved out by the two jets in the common dusty envelope surrounding the system.

In this letter we present high contrast PDI observations of Z\,CMa at $H$ and $K_{s}$
bands with NaCo. Our results show the excellent high contrast capabilities of this
instrument after its implementation from VLT/UT4 to VLT/UT1. We find
new polarized features down to an inner working angle of $\sim0\farcs12$,
and resolve the two components of Z\,CMa in the two bands showing that the
primary is driving a new outburst.
Our results allow us to describe the complex inner environment
of this young system with unprecedented detail.

\section{Observations}

We observed Z\,CMa and the comparison star HD\,52841 with NaCo
\citep{2003SPIE.4841..944L, 2003SPIE.4839..140R} on January 19th, 2015, right after its
commissioning at the VLT/UT1. The seeing was mostly stable (median $\sim0\farcs87$)
during the observations with a few short episodes of bad (above $1\farcs10$) values
and a median coherence time of $\tau_{0} = 3.5$ ms.
Z\,CMa was very bright ($m_{H}\le6$) and the Adaptive Optics (AO) system delivered
near diffraction limited images.

Our observations were obtained with the NIR detector with the $H$ and $K_s$ filters. The S27
camera ($27.05\pm0.1\,\mathrm{mas\,px^{-1}}$) was used in cube mode with the read-out set to
Double\_RdRstRd and the detector mode set to HighDynamic. We used individual exposure
times of 0.15~s and 0.5~s for Z\,CMa, and 1~s for HD\,52841. The frame loss rate in the short
exposure images was minimized by windowing the read-out region to $512\times520$ px. The
effect of bad pixels was reduced by dithering on the detector. HD\,52841 was all the time below
the non-linear NaCo regime ($10^{4}$ counts). For Z\,CMa, the primary is saturated in all our
images and several pixels surrounding both stars have values above the linear regime.
We used NaCo in polarimetry mode combining a half-wave plate (HWP) with a Wollaston
prism. In this setup the HWP rotates the polarization plane of the incoming light and the
Wollaston splits the light beam into two images with orthogonal polarization states
that are projected in different regions of the detector. A field mask avoids beam-overlapping. 
These two images are hereafter named as $I_\mathrm{o}$ (for the ordinary beam), and
$I_\mathrm{e}$ (for the extra-ordinary beam). We took images with the HWP rotated by
$0\degr, -22.5\degr, -45\degr$ and $-67.5\degr$ to reconstruct the Stokes parameters of
the incoming light. The observations are listed in Table~\ref{tab:tab1}.

\begin{table}[t]
		\center
				\caption{Observing log of the observations. Bad seeing frames do not
				contribute to the total exposure time listed below. All observations were
				taken during 19/01/2015.}
			\begin{tabular}{ c  c  c  c  c  }
\hline\hline
Target			&	Band	&		DIT		&	Tot. Exp Time	&	airmass	 \\
				&			&		[s]		&          [s]			&	[min, max]	 \\
\hline
Z\,CMa			&	$H$		&		0.15	&	957.9			&	1.03, 1.15	 \\
				&			&		0.5		&	166.0			&	1.04 ,1.05	 \\
				&	$K_s$	&		0.15	&	603.0			&	1.05 ,1.07	 \\
				&			&		0.5		&	176.0			&	1.07 ,1.10	 \\
HD\,52841		&	$H$		&		1		&	160.0			&	1.20 ,1.22    \\
				&	$K_s$	&		1		&	160.0			&	1.24, 1.26	 \\
\hline
			\end{tabular}
	\label{tab:tab1}
\end{table}

\section{Data reduction}

We used our own pipeline outlined in \citet{2011A&A...531A.102C, 2013A&amp;A...556A.123C}
to process the observations. Here we briefly describe the major aspects of the reduction. 
The images with the poorest AO correction (obtained under seeing $>1\farcs1$) were discarded in our analysis.
The rest of the images were dark subtracted, flat-field corrected and sky-background subtracted.
Hot and dead pixels were flagged. The images were first aligned with a cross-correlation algorithm.
We then refined the alignment up to an accuracy of 0.05 px with a minimization method. In short,
our algorithm finds the position that minimizes the standard deviation ($\sigma$) of the difference between
the two images to be aligned using cubic interpolation. The polarized information was extracted using the
double-difference approach \citep[e.g.][]{2009ApJ...701..804H, 2011A&A...531A.102C}. This way the
$I_\mathrm{o}$ and $I_\mathrm{e}$ images are added and subtracted to produce an intensity and polarized
image, respectively. When the HWP is at $0\degr$ the polarized image corresponds to the Stokes +Q image.
Similarly, when the HWP is at $22.5\degr, 45\degr$ and $67.5\degr$ the resulting polarized image corresponds
to Stokes +U, -Q, and -U, respectively. Instrumental polarization (IP) is usually corrected at this stage assuming
that the central star is unpolarized \citep[e.g.,][]{2014ApJ...781...87A}. Unfortunately, this is not the case for the two
stars in the Z\,CMa system \citep{1998A&A...334..969F}.
Additionally, the saturation of the primary
and the extended detector area around the two stars with pixel values well above the non-linear regime
of the NaCo detector complicate the polarimetric data reduction. We therefore measure the polarization degree
in a 1-px thick ($0\farcs027$) contour around the non-linear region for each individual image. 
This region contains contributions from the IP and the interstellar polarization and reaches a maximum of
$\sim2.5\%$\footnote{\citet{2011A&A...525A.130W} reports a maximum IP for NaCo at UT4 of $4\%$.}
in our observations. This quantity was subtracted from the individual images. To test for the effect of this correction
we also reduced our images using different amounts of correction ranging from $1\%$ to $4\%$. The non-linear
pixels are flagged and not considered in our polarimetric analysis. The images were median-combined
to construct the final Stokes parameters as $Q = (+Q - (-Q))/2$ and $U = (+U - (-U))/2$. The (linearly) polarized
intensity ($P_{I}$) is described by $P_{I} = \sqrt{Q^2 + U^2}$.  The polarization angle indicating the vibration
plane of the electric field is $P_{\theta} = \frac{1}{2}arctan (U/Q)$. The total intensity is computed by adding-up
the median-combined intensity images. Finally, the images were normalized to 1~s exposure time.

\section{Results}

The short-exposed images of Z\,CMa show the same structures as the long-exposure ones
while having smaller inner working angle. Therefore we focus our analysis on the 0.15~s images.
In what follows we adopt the most recent value of 930 pc for the distance to the source.

\begin{figure}[H]
  \centering \includegraphics[width=\columnwidth, trim = 10 35 15 35]{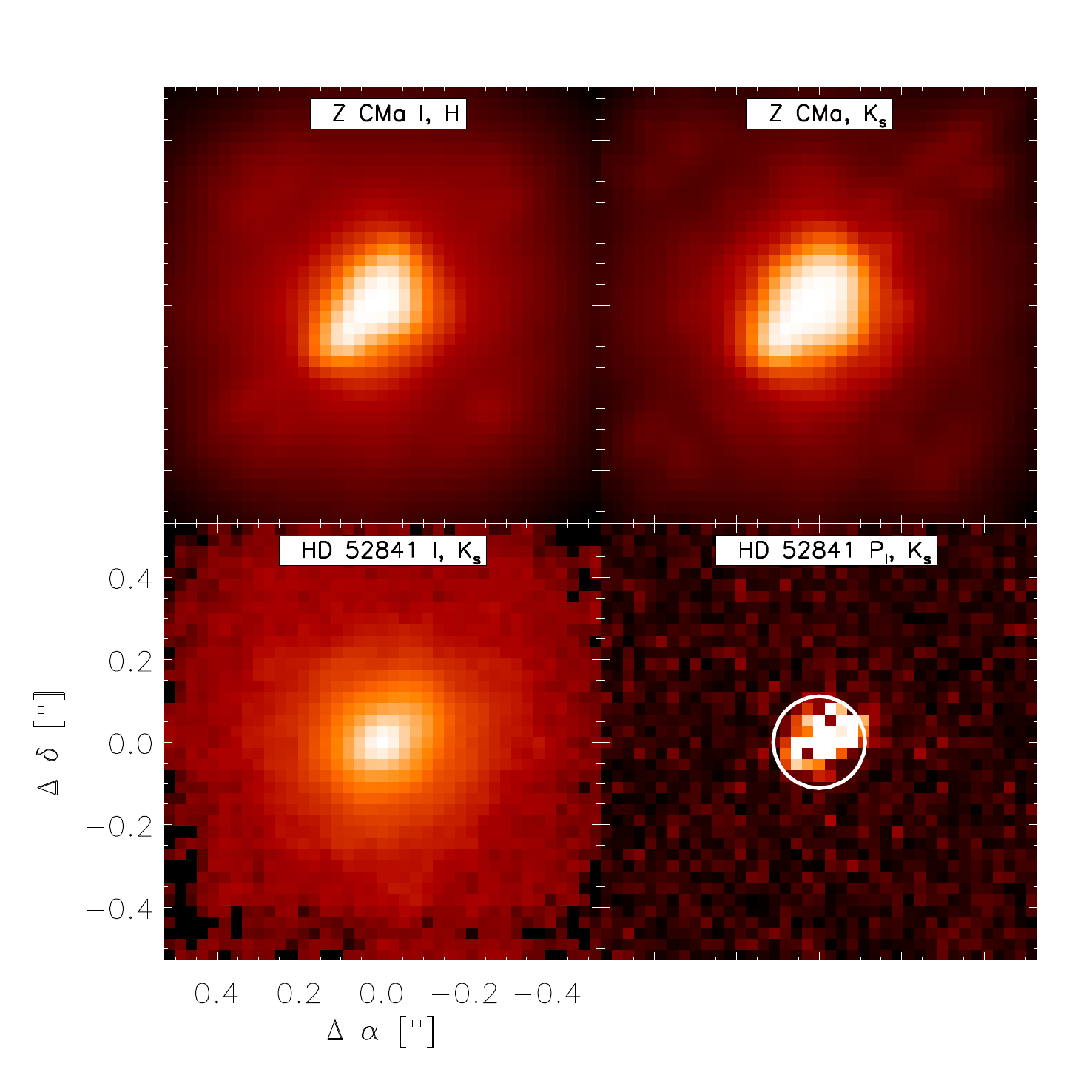}
  \caption{Top and bottom-left: Intensity images in logarithmic scale. AO artifacts are visible in the Z\,CMa images.
  HD\,52841 is slightly elongated in the SE-NW direction. Bottom-right: The $P_{I}$ image of HD\,52841 shows the noise
  that inevitably appears in the innermost $0\farcs1$ (indicated by the white circle) of the image as a result
  of imperfect speckle suppression.}
  
   \label{fig:fig1}
\end{figure}

The Z\,CMa binary is resolved in both $H$ and $K_s$ bands (Fig.~\ref{fig:fig1} upper panels). 
The primary component is saturated in both bands. To estimate the separation ($\rho$), position
angle ($PA$), and difference in flux ($\Delta m_{H,Ks}$) of the Z\,CMa binary system is not
straightforward because of the saturation of the primary and the small separation between both
stars. We circumvent this problem by fitting the primary's point spread function (PSF) wings and the
secondary's full PSF. This way we derive a separation of $\rho = 0\farcs112 \pm 0\farcs003$ and a
PA of $139.6\degr \pm 2.0\degr$. The brightness difference between the primary and secondary
is noticeably larger at $K_s$ than $H$ band, with $\Delta m_{H} = -1.1\pm0.2$ and $\Delta m_{Ks} \sol -1.9$,
similar to the values observed during an outburst driven by the primary \citep{2009ApJ...701..804H, 2010A&A...509L...7S}.

Comparing our results with those derived by \citet{2002ApJ...580L.167M} using data from 2001 we
detect an orbital motion of $\Delta PA = 10.6\degr\pm2.1\degr$ and $\Delta \rho = 0\farcs003\pm0\farcs003$
in 14 yr. Our results are in agreement with \citet{2002ApJ...580L.167M} who find a marginal change in $\rho$
but a significant change of $\Delta PA=8\fdg8\pm1\fdg5$ when analyzing previous observations covering a
11.2 yr baseline, deriving a $\Delta PA$ rate of $\sim0\fdg7\,\mathrm{yr^{-1}}$ for a circular orbit composed
of a $1M_{\sun}$ and a $5M_{\sun}$ stars. 

\begin{figure*}[t!]
  \centering \includegraphics[width=\textwidth, trim = 0 35 0 30]{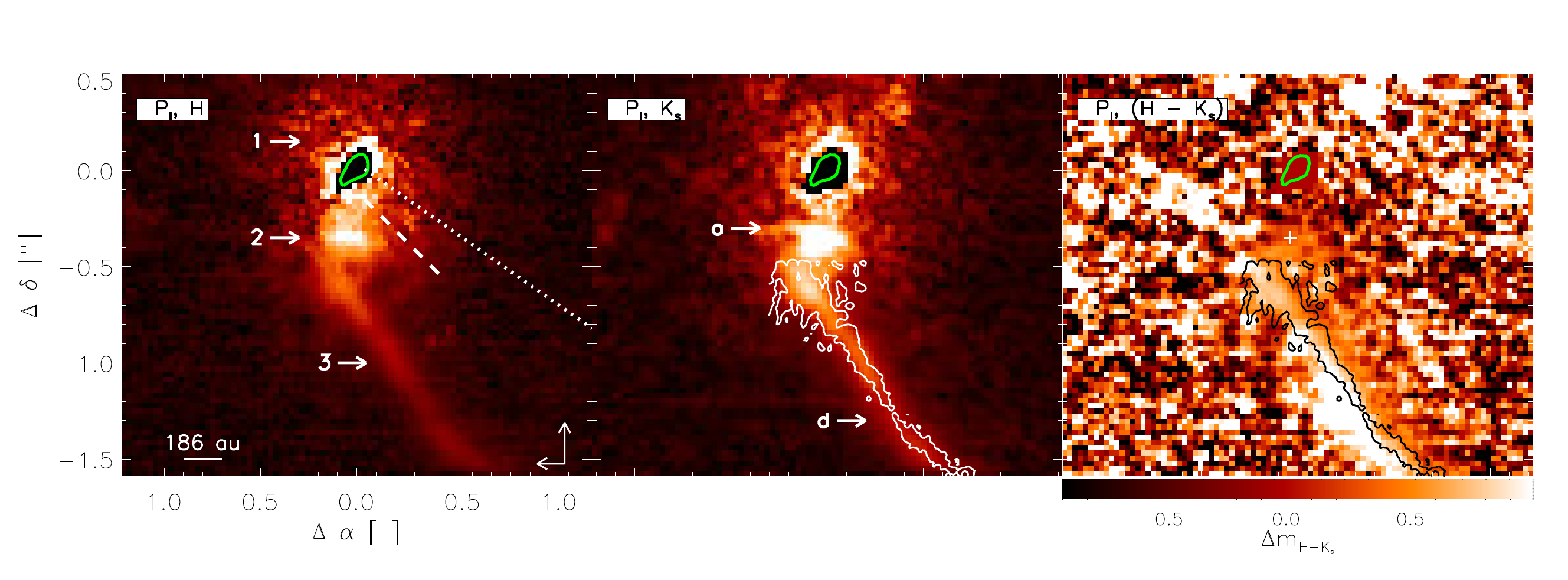}
  \caption{Right and center: $P_{I}$ images at $H$ and $K_s$ bands. Green contours indicate the position of the two stars. The two images are plot
  in the same scale, using square root scaling. The pixels falling in the non-linear regime are set to zero. Left: the dotted and dashed lines
  indicate the position of large and micro jets, respectively \citep{2010ApJ...720L.119W}. Arrows ``1'', ``2'' and ``3'' indicate the polarized halo,
  the bright clump, and the sharp extended feature, respectively. Center: Arrows ``a'' and ``d'' shows an artifact and a local decrement in polarized flux,
  respectively. The white line contours the Keck $J$ band image presented in \citet{2002ApJ...580L.167M}.
  Right: Relative polarized color computed as [$H$] - [$K_s$]. The white cross indicates the brightest
  point of feature ``2'' at $K_s$ band.} \label{fig:fig2}
\end{figure*}

Our polarized images (Fig.~\ref{fig:fig2}) show three complex features down to the inner
$\sim0\farcs12$ from the stars in the two bands. First, we detect an approximately circular
halo around the two stars (labeled as ``1'' in Fig.~\ref{fig:fig2}, left panel). The detector window
trims the upper part of the image, where the halo is detected at signal to noise (S/N) of $\sim8$,
suggesting that it extends beyond that limit. In the east-west direction the halo is detected up to
$\sim 0\farcs8$ ($\sim744$ au) from Z\,CMa. Second, we detect a bright, irregularly shaped polarized
clump at $\sim0\farcs3$ ($\sim279$ au projected distance) south from Z\,CMa (labeled as ``2''). This
clump is elongated in the east-west direction and it is separated by roughly 
$0\farcs15$ from the ``k1'' feature described by
\citet{2010ApJ...720L.119W}. Third, we re-detect with high S/N ($\sim30$) the sharp extended feature
(labeled as ``3'') previously observed by \citet{2002ApJ...580L.167M} with Keck at $J$ band (shown in
white contours in Fig.~\ref{fig:fig2}, central panel). This feature extends down to 0\farcs4 and appears
spatially connected with the clump, and is likely to extend south beyond our field of view because of its
high S/N ($\sim$28 at 1\farcs5 south). The relative polarized color (Fig.~\ref{fig:fig2}) shows that, in scattered light,
feature ``3'' is remarkably redder when compared to the neutral ($\Delta m_{H-K_s}\sim0$) polarized halo or with the
bright clump ($\Delta m_{H-K_s}\sim0.4$). Fig.~\ref{fig:fig2} also suggests that feature ``3'' is more extended
in the east-west direction in $K_s$ band than in $H$ band, specially towards the south direction. In both
bands the polarized flux along the feature decreases with projected distance to the star, but not uniformly.
There is a local decrement in $P_{I}$ at $\sim-1\farcs3$ south (see Fig.~\ref{fig:fig2}, central panel), also
observed in intensity at $J$ band in the Keck images. We find the same complex structures when
processing the observations using different amounts of IP correction, therefore concluding that they are not artifacts.

\begin{figure*}[t!]
  \centering \includegraphics[width=\textwidth, trim = 30 40 10 40]{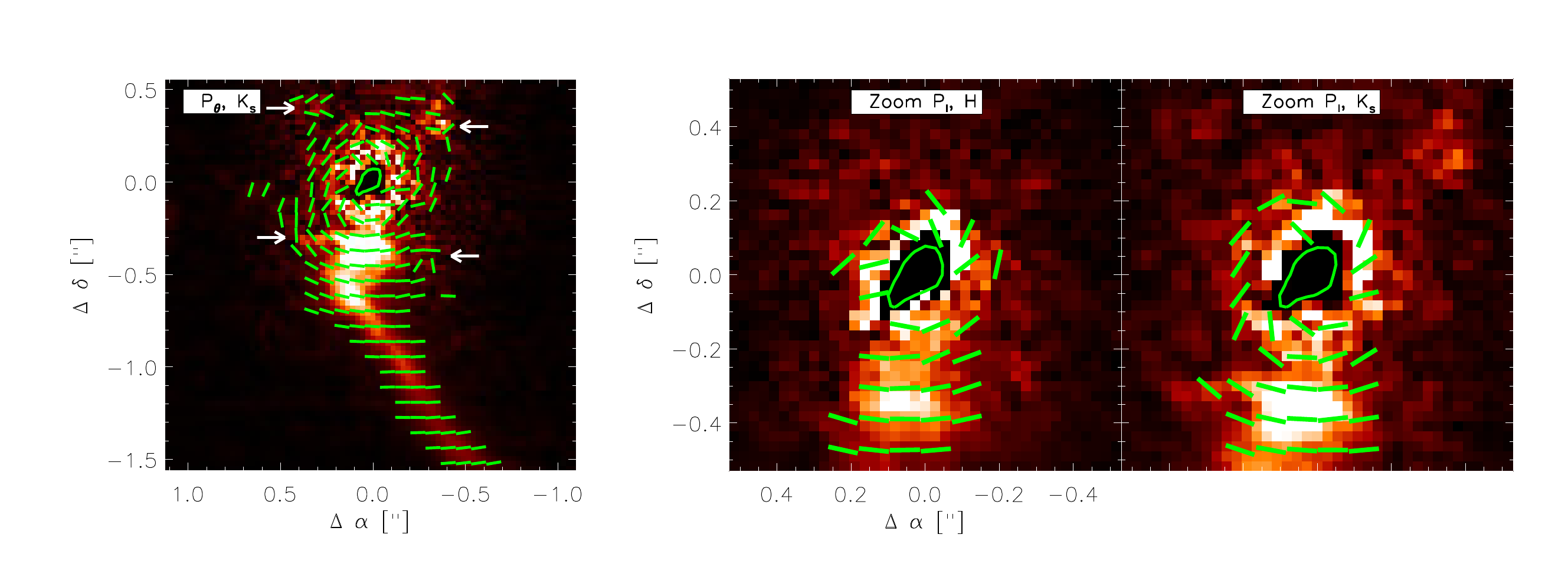}
  \caption{Left: $P_{\theta}$ over polarized intensity at $K_s$-band. The length of the vectors is arbitrary. The image is plot
  in lineal scale, stretching the scale to highlight faint features. AO artifacts are indicated by the arrows.} \label{fig:fig3}
\end{figure*}

The polarization angle $P_{\theta}$ at $K_s$ band is shown in Fig.~\ref{fig:fig3}, left panel.
The vectors are plotted only in regions of the image with S/N $>10$ averaging over a 3-px width
($\sim1.2$ FWHM at $K_s$ band) squared box. Overall the vectors follow an axi-symmetric pattern
around the Z\,CMa binary system and they show the same orientation along feature ``3''. The white arrows indicate polarized artifacts
created by the AO system. In these regions $P_{\theta}$ differs from the rest of the image, showing
a non axy-symmetrical orientation. An extension towards the East seems to depart from the polarized
clump (see Fig.~\ref{fig:fig2}, central panel). This feature is more likely an artifact, since it is located very
close to one AO artifact, and it does not appear in any of the other images. A zoom around the stars is
shown in the central and right panels of Fig.~\ref{fig:fig3}. In these plots we only show $P_{\theta}$ where
the $S/N>30$.

Interestingly, there is a polarized filament connecting the clump (feature ``2'') with the innermost regions of Z\,CMa.
The bright pixels around the non-linear (set to 0) pixels are likely to have a contribution from the inherent noise
observed in the inner $0\farcs1$ around the stars (see the $P_{I}$ image of the comparison star at the bottom
right panel in Fig.~\ref{fig:fig1}). However, the two stars and their respective circumstellar disks (and cocoon around
the primary) are known to be polarized \citep{1998A&A...334..969F}, and a large contribution from them
and the polarized halo can explain the quasi-axisymmetric distribution of $P_{\theta}$, specially at $K_s$ band.

\section{Discussion and Conclusions}

Our polarized observations show the complex and rich environment of Z\,CMa.
The nature of feature ``3'' remains unclear. It is not spatially close to the two jets detected by
\citet{2010ApJ...720L.119W}, and to date there are no detections of jets in continuum (while
there are several detections in emission lines). Particularly intriguing is the apparent change in direction
it follows before connecting to the bright clump, and the morphological similarities 
with broad cylindrical cavities recently observed in CO at late stages of star formation
\citep[][]{2005ApJ...624..841L, 2013ApJ...774...39A} and at $10\mu$m around massive embedded stars \citep{2014ApJ...796...74L}.
Although it is difficult to imagine how feature ``3'' can produce scattered light at $\le1400$ au (projected
distance) from the stars, this feature could be explained by a large filament of dust with different scattering
properties than its surrounding material. Alternatively, it could be created by light scattered off the walls 
of a cavity created by the interaction of a molecular outflow with the surrounding material.
We find no change in the shape of that feature when comparing with the Keck images taken
14 years ago (Fig.~\ref{fig:fig2}, central panel). Given the youth of Z\,CMa \citep[$\le3\,10^{5}$ yr, ][]{2004MNRAS.349.1516V}
and the size of the polarized clump (feature ``2''), a tantalizing explanation for this clump could be to relate it to an
embedded, very young star. High resolution images at mid infrared wavelengths can help to disentangle the nature
of this clump by providing evidence of a mid-IR excess (or lack thereof). 

We detect orbital motion on the binary system of Z\,CMa. Assuming a circular orbit composed of a
$1M_{\sun}$ and a $5M_{\sun}$ stars and comparing with previous measurements, our data suggest
an almost face-on orbit with a measurable orbital motion of $\Delta PA = 0\fdg7\,\mathrm{yr}^{-1}$,
in agreement with the values derived by \citet{2002ApJ...580L.167M}. The derived relative photometry
is consistent with a new outburst from the system.

The 1 and 3mm continuum images obtained by \citet{2009A&A...497..117A} with IRAM-PdBI show a partially
resolved structure elongated in the N-S direction that could be related to the observations here
presented. In any case, images at (sub-) mm wavelengths with much higher spatial resolution are urgently needed
to study the cold-dust counterpart of our observations to complete our picture of Z\,CMa and to
understand the earliest evolutionary stages of the binary systems.
Finally, our results show that NaCo's imaging polarimetric capabilities are at its best,
allowing us to probe the innermost regions of Z\,CMa


\begin{acknowledgements}
	We are grateful to George Hau and to the ESO staff for their help during the observations,
	and to J. Monnier for sharing the Keck $J$ band images. 
	This research was funded by the Millennium Science Initiative, Chilean Ministry of Economy, Nucleus
	RC130007. HC and CC acknowledge support from ALMA/CONICYT (grants 31100025 and 31130027).
	SP and SC acknowledge financial support provided by FONDECYT grants 3140601 and 1130949.
	CC and MRS acknowledge support from CONICYT-FONDECYT grant 3140592 and FONDECYT grant 1141269, respectively.
	LC was supported by ALMA-CONICYT and CONICYT-FONDECYT 31120009 and 1140109.
\end{acknowledgements}

\bibliographystyle{aa.bst}	
\bibliography{zcma.bib}		

\begin{thebibliography}{27}
\expandafter\ifx\csname natexlab\endcsname\relax\def\natexlab#1{#1}\fi

\bibitem[{{Alonso-Albi} {et~al.}(2009){Alonso-Albi}, {Fuente}, {Bachiller},
  {Neri}, {Planesas}, {Testi}, {Bern{\'e}}, \& {Joblin}}]{2009A&A...497..117A}
{Alonso-Albi}, T., {Fuente}, A., {Bachiller}, R., {et~al.} 2009, \aap, 497, 117

\bibitem[{{Arce} {et~al.}(2013){Arce}, {Mardones}, {Corder}, {Garay},
  {Noriega-Crespo}, \& {Raga}}]{2013ApJ...774...39A}
{Arce}, H.~G., {Mardones}, D., {Corder}, S.~A., {et~al.} 2013, \apj, 774, 39

\bibitem[{{Avenhaus} {et~al.}(2014){Avenhaus}, {Quanz}, {Schmid}, {Meyer},
  {Garufi}, {Wolf}, \& {Dominik}}]{2014ApJ...781...87A}
{Avenhaus}, H., {Quanz}, S.~P., {Schmid}, H.~M., {et~al.} 2014, \apj, 781, 87

\bibitem[{{Beckwith} \& {Sargent}(1991)}]{1991ApJ...381..250B}
{Beckwith}, S.~V.~W. \& {Sargent}, A.~I. 1991, \apj, 381, 250

\bibitem[{{Benisty} {et~al.}(2010){Benisty}, {Malbet}, {Dougados}, {Natta}, {Le
  Bouquin}, {Massi}, {Bonnefoy}, {Bouvier}, {Chauvin}, {Chesneau}, {Garcia},
  {Grankin}, {Isella}, {Ratzka}, {Tatulli}, {Testi}, {Weigelt}, \&
  {Whelan}}]{2010A&amp;A...517L...3B}
{Benisty}, M., {Malbet}, F., {Dougados}, C., {et~al.} 2010, \aap, 517, L3

\bibitem[{{Canovas} {et~al.}(2013){Canovas}, {M{\'e}nard}, {Hales},
  {Jord{\'a}n}, {Schreiber}, {Casassus}, {Gledhill}, \&
  {Pinte}}]{2013A&amp;A...556A.123C}
{Canovas}, H., {M{\'e}nard}, F., {Hales}, A., {et~al.} 2013, \aap, 556, A123

\bibitem[{{Canovas} {et~al.}(2012){Canovas}, {Min}, {Jeffers}, {Rodenhuis}, \&
  {Keller}}]{2012A&amp;A...543A..70C}
{Canovas}, H., {Min}, M., {Jeffers}, S.~V., {Rodenhuis}, M., \& {Keller}, C.~U.
  2012, \aap, 543, A70

\bibitem[{{Canovas} {et~al.}(2011){Canovas}, {Rodenhuis}, {Jeffers}, {Min}, \&
  {Keller}}]{2011A&A...531A.102C}
{Canovas}, H., {Rodenhuis}, M., {Jeffers}, S.~V., {Min}, M., \& {Keller}, C.~U.
  2011, \aap, 531, A102+

\bibitem[{{Clari{\'a}}(1974)}]{1974A&A....37..229C}
{Clari{\'a}}, J.~J. 1974, \aap, 37, 229

\bibitem[{{Fischer} {et~al.}(1998){Fischer}, {Stecklum}, \&
  {Leinert}}]{1998A&A...334..969F}
{Fischer}, O., {Stecklum}, B., \& {Leinert}, C. 1998, \aap, 334, 969

\bibitem[{{Hinkley} {et~al.}(2013){Hinkley}, {Hillenbrand}, {Oppenheimer},
  {Rice}, {Pueyo}, {Vasisht}, {Zimmerman}, {Kraus}, {Ireland}, {Brenner},
  {Beichman}, {Dekany}, {Roberts}, {Parry}, {Roberts}, {Crepp}, {Burruss},
  {Wallace}, {Cady}, {Zhai}, {Shao}, {Lockhart}, {Soummer}, \&
  {Sivaramakrishnan}}]{2013ApJ...763L...9H}
{Hinkley}, S., {Hillenbrand}, L., {Oppenheimer}, B.~R., {et~al.} 2013, \apjl,
  763, L9

\bibitem[{{Hinkley} {et~al.}(2009){Hinkley}, {Oppenheimer}, {Soummer},
  {Brenner}, {Graham}, {Perrin}, {Sivaramakrishnan}, {Lloyd}, {Roberts}, \&
  {Kuhn}}]{2009ApJ...701..804H}
{Hinkley}, S., {Oppenheimer}, B.~R., {Soummer}, R., {et~al.} 2009, \apj, 701,
  804

\bibitem[{{Kaltcheva} \& {Hilditch}(2000)}]{2000MNRAS.312..753K}
{Kaltcheva}, N.~T. \& {Hilditch}, R.~W. 2000, \mnras, 312, 753

\bibitem[{{Koresko} {et~al.}(1991){Koresko}, {Beckwith}, {Ghez}, {Matthews}, \&
  {Neugebauer}}]{1991AJ....102.2073K}
{Koresko}, C.~D., {Beckwith}, S.~V.~W., {Ghez}, A.~M., {Matthews}, K., \&
  {Neugebauer}, G. 1991, \aj, 102, 2073

\bibitem[{{Kraus} {et~al.}(2012){Kraus}, {Ireland}, {Hillenbrand}, \&
  {Martinache}}]{2012ApJ...745...19K}
{Kraus}, A.~L., {Ireland}, M.~J., {Hillenbrand}, L.~A., \& {Martinache}, F.
  2012, \apj, 745, 19

\bibitem[{{Lee} \& {Ho}(2005)}]{2005ApJ...624..841L}
{Lee}, C.-F. \& {Ho}, P.~T.~P. 2005, \apj, 624, 841

\bibitem[{{Lenzen} {et~al.}(2003){Lenzen}, {Hartung}, {Brandner}, {Finger},
  {Hubin}, {Lacombe}, {Lagrange}, {Lehnert}, {Moorwood}, \&
  {Mouillet}}]{2003SPIE.4841..944L}
{Lenzen}, R., {Hartung}, M., {Brandner}, W., {et~al.} 2003, Proc. SPIE, 4841,
  944

\bibitem[{{Li} {et~al.}(2014){Li}, {Mari{\~n}as}, \&
  {Telesco}}]{2014ApJ...796...74L}
{Li}, D., {Mari{\~n}as}, N., \& {Telesco}, C.~M. 2014, \apj, 796, 74

\bibitem[{{Millan-Gabet} \& {Monnier}(2002)}]{2002ApJ...580L.167M}
{Millan-Gabet}, R. \& {Monnier}, J.~D. 2002, \apjl, 580, L167

\bibitem[{{Poetzel} {et~al.}(1989){Poetzel}, {Mundt}, \&
  {Ray}}]{1989A&A...224L..13P}
{Poetzel}, R., {Mundt}, R., \& {Ray}, T.~P. 1989, \aap, 224, L13

\bibitem[{{Rousset} {et~al.}(2003){Rousset}, {Lacombe}, {Puget}, {Hubin},
  {Gendron}, {Fusco}, {Arsenault}, {Charton}, {Feautrier}, {Gigan}, {Kern},
  {Lagrange}, {Madec}, {Mouillet}, {Rabaud}, {Rabou}, {Stadler}, \&
  {Zins}}]{2003SPIE.4839..140R}
{Rousset}, G., {Lacombe}, F., {Puget}, P., {et~al.} 2003, Proc. SPIE, 4839, 140

\bibitem[{{Szeifert} {et~al.}(2010){Szeifert}, {Hubrig}, {Sch{\"o}ller},
  {Sch{\"u}tz}, {Stelzer}, \& {Mikul{\'a}{\v s}ek}}]{2010A&A...509L...7S}
{Szeifert}, T., {Hubrig}, S., {Sch{\"o}ller}, M., {et~al.} 2010, \aap, 509, L7+

\bibitem[{{van den Ancker} {et~al.}(2004){van den Ancker}, {Blondel}, {Tjin A
  Djie}, {Grankin}, {Ezhkova}, {Shevchenko}, {Guenther}, \&
  {Acke}}]{2004MNRAS.349.1516V}
{van den Ancker}, M.~E., {Blondel}, P.~F.~C., {Tjin A Djie}, H.~R.~E., {et~al.}
  2004, \mnras, 349, 1516

\bibitem[{{Whelan} {et~al.}(2010){Whelan}, {Dougados}, {Perrin}, {Bonnefoy},
  {Bains}, {Redman}, {Ray}, {Bouy}, {Benisty}, {Bouvier}, {Chauvin}, {Garcia},
  {Grankvin}, \& {Malbet}}]{2010ApJ...720L.119W}
{Whelan}, E.~T., {Dougados}, C., {Perrin}, M.~D., {et~al.} 2010, \apjl, 720,
  L119

\bibitem[{{Whitney} {et~al.}(1993){Whitney}, {Clayton}, {Schulte-Ladbeck},
  {Calvet}, {Hartmann}, \& {Kenyon}}]{1993ApJ...417..687W}
{Whitney}, B.~A., {Clayton}, G.~C., {Schulte-Ladbeck}, R.~E., {et~al.} 1993,
  \apj, 417, 687

\bibitem[{{Whitney} \& {Hartmann}(1993)}]{1993ApJ...402..605W}
{Whitney}, B.~A. \& {Hartmann}, L. 1993, \apj, 402, 605

\bibitem[{{Witzel} {et~al.}(2011){Witzel}, {Eckart}, {Buchholz}, {Zamaninasab},
  {Lenzen}, {Sch{\"o}del}, {Araujo}, {Sabha}, {Bremer}, {Karas}, {Straubmeier},
  \& {Muzic}}]{2011A&A...525A.130W}
{Witzel}, G., {Eckart}, A., {Buchholz}, R.~M., {et~al.} 2011, \aap, 525, A130

\end{thebibliography}

\clearpage
\newpage


\end{document}